\documentclass[12pt]{article}
\usepackage{amsmath}
\usepackage{graphicx,psfrag,epsf}
\usepackage{enumerate}
\usepackage{natbib}
\usepackage{amsfonts}
\usepackage{mathrsfs}
\usepackage{caption}
\newcommand{\blind}{0}

\def\half{\hbox{$1\over2$}}

\def\half{\hbox{$1\over2$}}

\begin{document}

\def\spacingset#1{\renewcommand{\baselinestretch}%
{#1}\small\normalsize} \spacingset{1}


\if0\blind
{
    \title{\bf A Latent Slice Sampling Algorithm}
    \author{Yanxin Li \thanks{
    For correspondence, contact: yanxinli@utexas.edu.}\hspace{.2cm}
    \\
    \normalsize{Department of Statistics and Data Sciences, University of Texas at Austin}\\ \\
    and \\ \\
    Stephen G. Walker \\
    \normalsize{Department of Mathematics, University of Texas at Austin}}
    \date{}
    \maketitle
} \fi

\if1\blind
{
  \bigskip
  \bigskip
  \bigskip
  \begin{center}
    {\LARGE\bf Title}
\end{center}
  \medskip
} \fi

\bigskip
\begin{abstract}
In this paper we introduce a new sampling algorithm which has the potential to be adopted as a universal replacement to the Metropolis--Hastings algorithm. It is related to the slice sampler, and motivated by an algorithm which is applicable to discrete probability distributions 
which obviates the need for a proposal distribution, in that is has no accept/reject component. 
This paper looks at the continuous counterpart. A latent variable combined with a slice sampler and a shrinkage procedure applied to uniform density functions creates a highly efficient sampler which can generate random variables from very high dimensional distributions as a single block.
\end{abstract}

\noindent%
{\it Keywords:}  High dimensional density; Markov chain Monte Carlo;  Shrinkage procedure; Uniform random variables.

\section{Introduction}
\label{sec:intro}

The original, and still one of the most popular sampling methods is the Metropolis--Hastings algorithm (Metropolis et al, 1953; Hastings, 1970). It generates a Markov sample which has a target density as the stationary density. One well known drawback of the algorithm is that the sampler can get stuck if the proposal density is not well set. In this case the Markov chain can linger at a particular value before, if ever, moving. Hence, if the density of interest can be sampled directly, via a rejection algorithm, or a Gibbs sampler, if appropriate, then these would be the methods of choice.   

The question discussed in this paper is whether we can always avoid a Metropolis--Hastings algorithm without compromising efficiency. When the sample space is discrete, say $\Omega=\{0,1,2,\ldots\}$,  the sampler presented in Walker (2014) is one such possibility. One of the key ideas behind the Metropolis--Hastings algorithm is the transition density $p(y\mid x)$, defined for all $x,y\in\Omega$, satisfying
\begin{equation}\label{reverse}
p(y\mid x)\,\pi(x)=p(x\mid y)\,\pi(y)
\end{equation}
where $\pi$ is the target density. The Metropolis--Hastings algorithm has transition density
$p(y\mid x)=\alpha(x,y)\,q(y\mid x)+(1-r(x))\,{\bf 1}(y=x),$
where $q(y\mid x)$ is a proposal density, to be chosen, 
$$\alpha(x,y)=\min\left\{1,\frac{\pi(y)\,q(x\mid y)}{\pi(x)\,q(y\mid x)}\right\},$$
and $r(x)=\int \alpha(x,y)\,q(y\mid x)\,d y$. It is easily seen that this $p(\cdot\mid\cdot)$ satisfies equation (\ref{reverse}).

An alternative $p(\cdot\mid\cdot)$ satisfying equation (\ref{reverse}) and proposed in Walker (2014) is given by
\begin{equation}\label{discrete}
p(y\mid x)=\frac{\pi(y)}{k}\,\sum_{l=\max(y,\, x)}^{\min(y+k-1, \,x+k-1)} \frac{1}{\sum_{z=\max(0, \,l-k+1)}^l \pi(z)},
\end{equation}
where $|y-x|<k$, and $k$ is to be chosen. However, the choice of $k$ is  easy to set; as large as possible while computations required to sample $p(y\mid x)$ remain time feasible. 
So note that with this transition density there is no possibility for the sampler to get stuck and neither is there an accept/reject component. Note also that $\pi$ only needs to be known up to a normalizing constant, a strong requirement in any sampler, as often, in many applications, the target density is only specified up to an unknown normalizing constant. Finally, note that (\ref{discrete}) is easy to sample.
A multivariate version of (\ref{discrete}) is easy to establish and has been applied to a certain class of optimization problem in Ekin et al. (2020).

The aim in the present paper is to find a continuous counterpart to (\ref{discrete}). In fact a suitable transition density is not difficult to write down as a direct analog of (\ref{discrete});
\begin{equation}\label{contin}
p(y\mid x)=\frac{\pi(y)}{k}\int_{l=\max(y,\, x)}^{\min(y+k,\, x+k)} \frac{dl}{\int_{z = l - k}^l \pi(z) d z},
\end{equation}
where here we have $\Omega=(-\infty,\infty)$ and $|y-x|\leq k$. Just as (\ref{discrete}) can be seen as a Gibbs sampler, so  can (\ref{contin}). To see this, consider the joint density function
\begin{equation}\label{joint}
p(y,l)=\pi(y)\,\frac{{\bf 1}(y<l<y+k)}{k},
\end{equation}
so clearly $\pi(y)$ is the required marginal density. Then (\ref{contin}) is given by
$p(y\mid x)=\int p(y\mid l)\,p(l\mid x)\,dl,$
where $p(l\mid x)$ is uniform on the interval $(x,x+k)$. Further, (\ref{contin}) also satisfies equation (\ref{reverse}).
The only outstanding question is how to sample (\ref{contin}), which is the main focus of the paper. Indeed, we show that sampling (\ref{contin}) can be done efficiently using only uniform random variables and only requires  the implementation of an adaptive rejection algorithm, which works extremely fast.


In section 2 the aim is to show how to sample from $p(y\mid x)$ given by (\ref{contin}) but with necessary extensions involving making $k$ random. This also requires some further latent variable; specifically a ``slice'' variable, similar in spirit to Besag and Green (1993), Damien et al (1999) and Neal (2003). Slice sampling, as it has become known, is a popular approach to sampling complex densities usually within a Gibbs sampling framework.
In fact slice samplers have good convergence properties; Robert and Rosenthal (1999) show that slice samplers are nearly always geometrically ergodic while Mira and Tierney (2002) provide sufficient conditions for a slice sampler to be uniformly ergodic.
Recent uses of Neal's approach include the elliptical slice sampler, see Murray et al (2010), and the generalized elliptical slice sampler, see Nishihara et al (2014), and factor slice sampling, see Tibbits et al (2014). Once the slice variable has been incorporated within (\ref{contin}), it is then possible to compare the new sampler with Neal's slice sampler. Indeed, as it stands with $k$ fixed,  it is precisely a version of Neal's algorithm. Both us and Neal extend from this fixed $k$, but in different directions. Neal adopts the reversible framework while we adopt a random $k$ approach and use the framework established by the joint density (\ref{contin}). This allows us to maintain a Gibbs sampling framework while avoiding a tricky detailed balance constraint. 
We make a direct comparison with Neal's slice sampler in section 3. Numerous illustrations are presented in section 4 and section 5 concludes with a brief description and a full layout of the algorithm for arbitrary multivariate distribution.

\section{Latent slice sampler} We first describe the algorithm in one dimension and later detail the extension to multi--dimensions. To develop the joint density (\ref{joint}), we make it more flexible by allowing $k$ to be a random variable, which we will now refer to as $s$, and assign $s$ to have density $p(s)$, to be chosen, and allow for $l$ to be in the interval $(y-s/2,y+s/2)$. 
Hence, the joint density of interest becomes
\begin{equation}\label{jjoint}
p(y,s,l)=\pi(y)\,p(s)\,\frac{{\bf 1}\big(y-s/2<l<y+s/2\big)}{s}.
\end{equation} 
A key aspect of the innovation in the sampler is on dispaly here; we have introduced a $y$ term outside of $\pi(
y)$ term without altering the correct marginal. 
So the marginal density of $y$ is $\pi(y)$ and the marginal density of $s$ is $p(s)$. A Gibbs sampler based directly on (\ref{jjoint}) would be difficult to implement as it is not possible to sample from $\pi(y)$; or rather it is assumed not to be able to do so. In such cases, a slice sampler can be utilized. By introducing a slice variable $w$, the joint density then becomes
\begin{equation}\label{jjjoint}
p(y,w,s,l)={\bf 1}\big(\pi(y) > w\big)\,p(s)\,\frac{{\bf 1}\big(y-s/2<l<y+s/2\big)}{s}.
\end{equation}
While this is more than used by Neal (2003), the extra component, i.e.
$$p(s)\,\frac{{\bf 1}\big(y-s/2<l<y+s/2\big)}{s}$$
is effectively providing the stochastic search engine for the set of $y$ for which $\pi(y)>w$. Such a  procedure was also required by Neal (2003) who used a search strategy while needing also to maintain a detailed balance criterion. On the other hand, we are free from some such constraints.
For us, this is greatly simplified, yet just as effective, by incorporating the search component into the joint density. This  means we do not have to implement a stepping out or a doubling procedure which is  a part of Neal's algorithm.

We implement a Gibbs sampler based on (\ref{jjjoint}). So $p(w,l\mid y,s)$ is easy to sample; being two conditionally independent uniform random variables. 
Further
\begin{equation}\label{cond}
p(s\mid y, w,l)\propto \frac{p(s)}{s}\,{\bf 1}\big(s>2|l-y|\big).
\end{equation}
This conditional density is also straightforward to sample; and throughout we take $p(s)\propto s\,e^{-\lambda s}$ for some $\lambda$, typically in order to provide a large variance. Finally,
$$p(y\mid w, s, l)\propto {\bf 1}\big(\pi(y)>w\big)\,{\bf 1}(l-s/2<y<l+s/2).$$
We sample this using an adaptive rejection sampler; it is also a shrinkage procedure as described in Neal (2003). Before describing the adaptive rejection sampler we present a simple illustration of the key aspects of the one step algorithm, starting with the current value $y_0$. 

\setlength\belowcaptionskip{-3ex}
\begin{figure}[!htbp]
\begin{center}
\includegraphics[height=5.6cm,width=14cm]{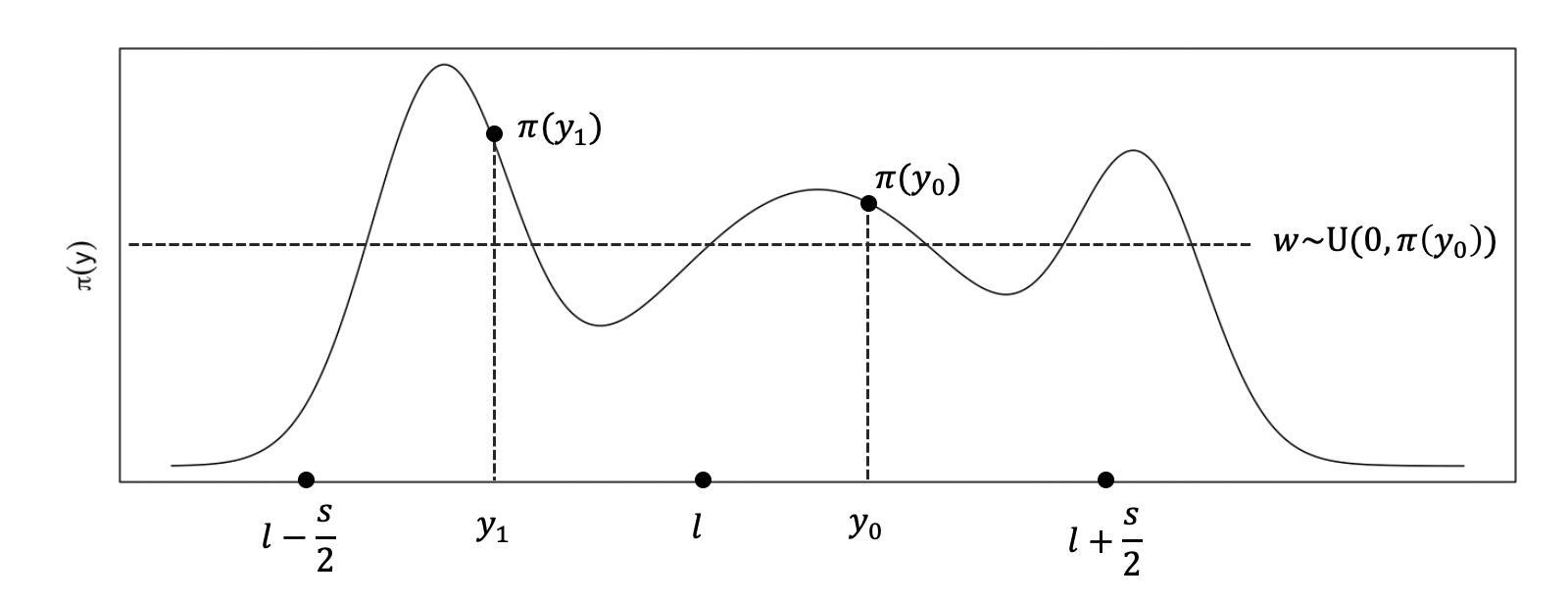}
\caption{Illustration of latent slice sampler}
\label{fplot}
\end{center}
\end{figure}

An illustration is provided in Fig.~\ref{fplot}. The current values of $y_0$, $w$ and $l$ are indicated. The illustration for this case gives a value of $s$ for which the relevant values of $l-s/2$ and $l+s/2$ are indicated. The proposed value of $y_1$ is sampled uniformly from $(l-s/2,l+s/2)$ and is accepted if $\pi(y_1)>w$, as shown in the graph. Rejected $y$ give information about the location of the interval $\pi(y)>w$ and this can be used to improve the proposal with the shrinkage procedure.
To generalize the setting we consider the adaptive rejection sampling of 
$$p(y)\propto {\bf 1}(y\in C)\,{\bf 1}(a<y<b),$$
where $C\cap (a,b)\neq \emptyset$ and $y_0\in C\cap (a,b)$. 
Let $a_1=a$ and $b_1=b$; at iteration $m$, starting at $m=1$,

\begin{itemize}
    \item[1.] Sample $y^*$ uniformly from $(a_m,b_m)$. 
    
    \item[2.] While $y^*\notin C$: if $y^*<y_{0}$ then $a_{m+1} \leftarrow \max\{a_m,y^*\}$ else $b_{m+1}\leftarrow \min\{b_m,y^*\}$ and $m\to m+1$.
    
    \item[3.] Repeat steps 1. and 2. until $y^*\in C$; then $y=y^*$.
    

\end{itemize}

\noindent
This works for reasons outlined in Neal (2003), and see also the discussion by  Walker in Neal's paper. The basic idea is that the sampling strategy resulting in $y=y^*$ conditional on $y_0$, and write this density as $p(y\mid y_0)$, satisfies detailed balance with respect to $p(y)$; i.e.
$$p(y\mid y_0)\,p(y)=p(y_0\mid y)\,p(y).$$
The obvious points here are that as $p(y)$ is uniform, one only need establish that
$p(y\mid y_0)=p(y_0\mid y)$ which is straightforward to understand.

\vspace{0.2in}
\noindent
{\sc Example 1.}
To see how efficient this sampling strategy is, we take the target for $y$ as a mixture of two normal densities with variances 1 and means -10 and +10, and with equal weights. That is,
$$\pi(y)=\half\,\mbox{N}(y\mid -10,1)+\half\,\mbox{N}(y\mid 10,1).$$ 

\setlength\belowcaptionskip{-3ex}
\begin{figure}[!htbp]
\begin{center}
\includegraphics[height=7cm,width=14cm]{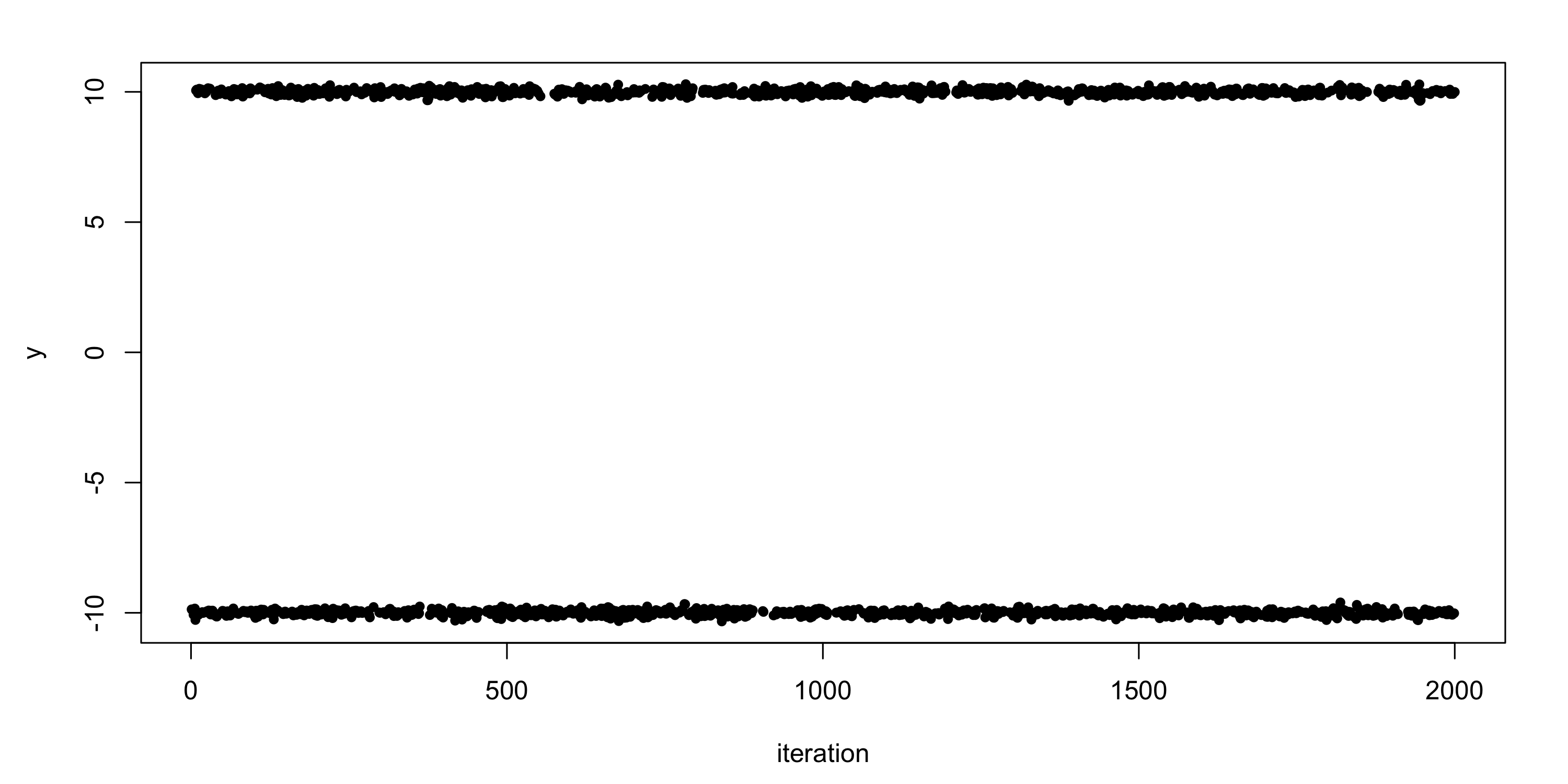}
\caption{Samples from latent slice algorithm from mixture of two normals}
 \label{fig1}
\end{center}
\end{figure}

We take $p(s)$ to be a gamma distribution with parameters shape equal to 2 and scale equal to 100, i.e., $p(s)\propto s\exp(-0.01 s)$, and generate 2,000 samples from the algorithm. The subsequent plot of the sampled $y$ is given in Fig.~\ref{fig1}. As can be seen, the mixing and accuracy of the samples is excellent. It should be noted that there are very few, if any, alternative algorithms using Markov chains, which could achieve this.

\subsection{Multivariate case} From the univariate case there is an easy way to set up a multivariate latent slice sampler when $y$ is a $d$--dimensional variable. We have the relevant joint density now as
$$p(y,w,s,l)={\bf 1}\big(\pi(y)>w\big)\,p(s)\prod_{j=1}^d \frac{{\bf 1}(l_j-s_j/2<y_j<l_j+s_j/2)}{s_j}.$$
So $w$ remains a one dimensional variable, but the other two; i.e. $s$ and $l$, are both $d$--dimensional.

The sampling strategy using a Gibbs sampler is an obvious extension to the one dimensional case. 
The conditional for $y$ is given by
$$p(y \mid w,s,l)\propto {\bf 1}\big(\pi(y)>w\big)\,\prod_{j=1}^d {\bf 1}(l_j-s_j/2<y_j<l_j+s_j/2).$$
This can also be sampled using the shrinkage procedure; writing $a_j=l_j-s_j/2$, $b_j=l_j+s_j/2$, $y_0=(y_{01},\ldots,y_{0d})$ as the current $y$, and $\{y:\pi(y)>w\}=C$, we sample proposal $y^*=(y_1^*,\ldots,y_d^*)$ from $\prod_{j=1}^d {\bf 1}(a_j<y_j<b_j)$ and accept $y=y^*$ if $y^*\in C$. Otherwise, do for all $j=1,\ldots,d$:
$$\mbox{if}\quad y^*_j<y_{0j}\quad\mbox{then}\quad a_j\leftarrow\max\{a_j,y^*_j\}\quad\mbox{else}\quad b_j\leftarrow\min\{b_j,y^*_j\}.$$

\begin{figure}[!htbp]
\begin{center}
\includegraphics[height=6.5cm,width=13cm]{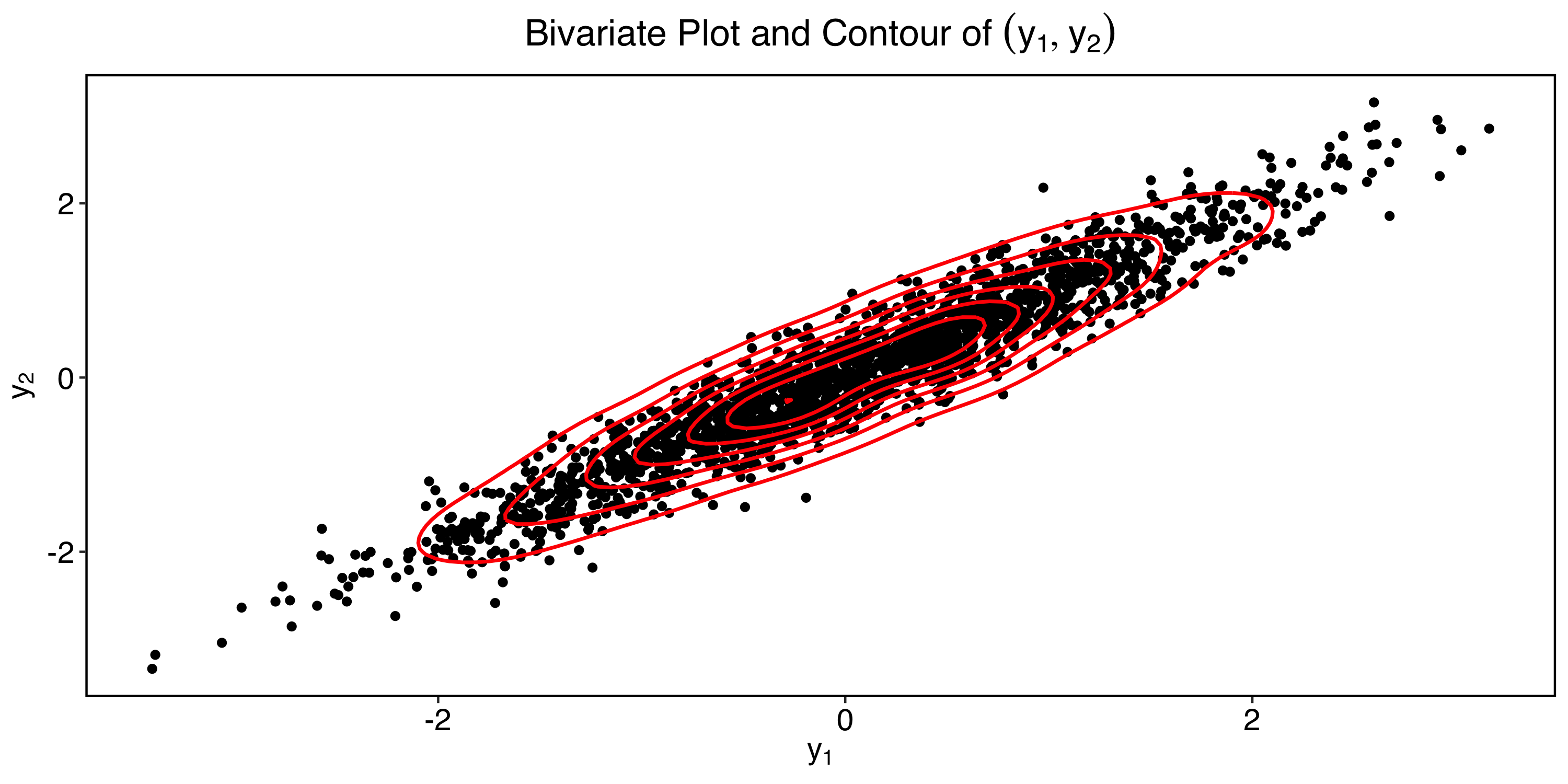}
\caption{Samples from latent slice algorithm from bivariate normal}
\label{fig2}
\end{center}
\end{figure}

\vspace{0.2in}
\noindent
{\sc Example 2}.
As an illustration we take $\pi(y)$ to be a bivariate normal density with a very high correlation; i.e. we take a mean of $(0,0)$ and a covariance matrix with unit variances and correlation $\rho=0.95$. It is known that slice sampling  algorithms can perform poorly when the variables are highly correlated; indeed, as stated in Tibbits et al (2014), ``It is particularly difficult to create an efficient sampler when there is strong dependence among the variables''. We take $p(s)$ to be independent gamma distributions with shape equal to 2 and scale equal to 10. The bivariate plot and contour of the $(y_1,y_2)$ from the output of the sampling algorithm is presented in Fig.~\ref{fig2}. As can be seen this has worked extremely well.

\setlength\belowcaptionskip{-3ex}

\begin{figure}[!htbp]
\begin{center}
\includegraphics[height=8cm,width=14cm]{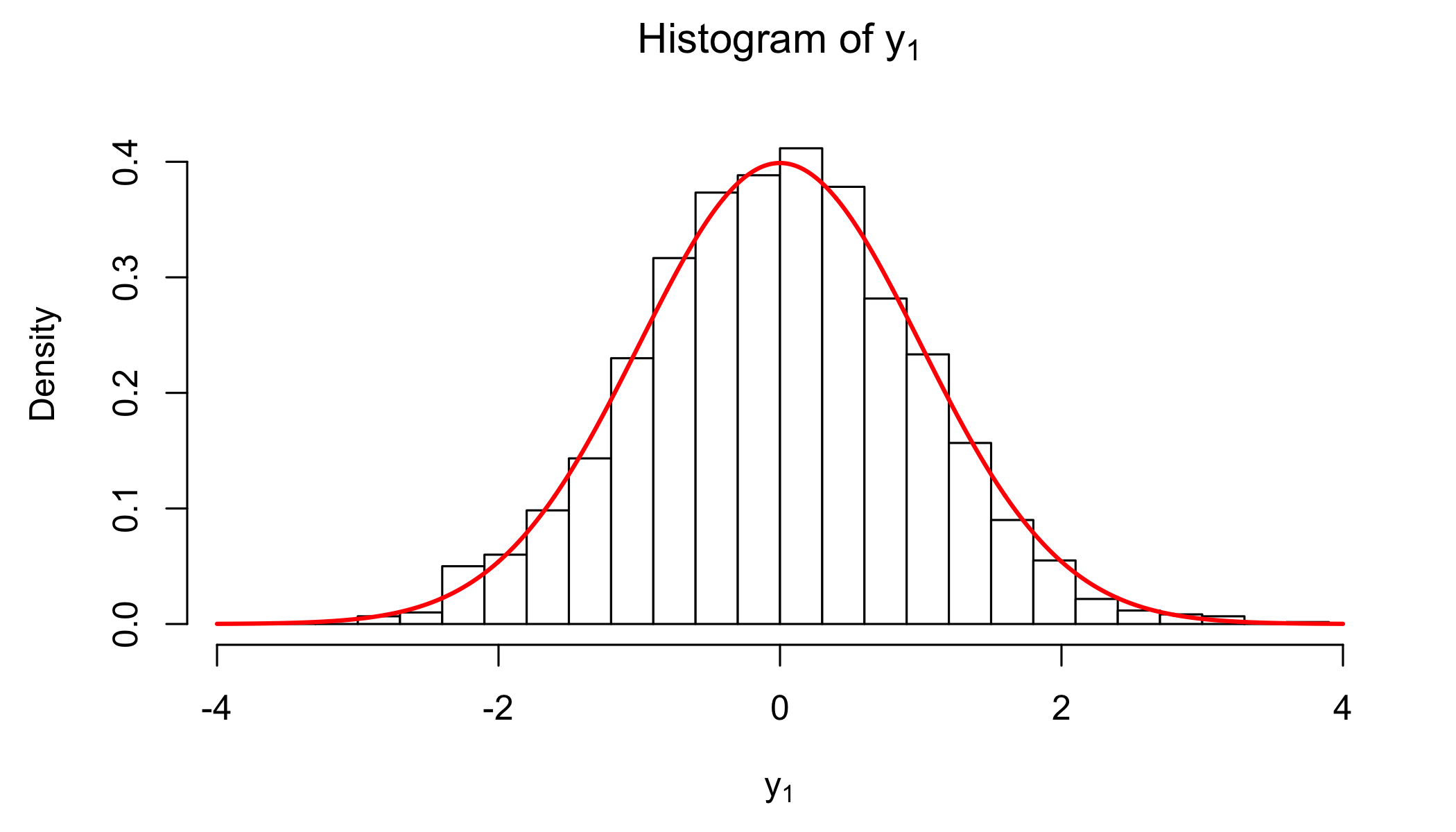}
\caption{Samples of $y_1$ from latent slice algorithm with 50 dimensional multivariate normal target density}
\label{figmult}
\end{center}
\end{figure}

\vspace{0.1in}
\noindent
{\sc Example 3}. Here we do a $d=50$ dimensional example with the target density 
$$\pi(y)\propto\exp\left\{-\half \sum_{j=1}^{d}y_j^2\right\}.$$
The code was written in R and 5000 samples of $y$ were collected. The time for execution was two seconds. We take the same $p(s)$ as that of Example 2. The samples of $y_1$ are presented as a histogram in Fig.~\ref{figmult} along with the standard normal density function for comparison.

\setlength\belowcaptionskip{-3ex}


\section{Comparison with slice sampling}

The algorithm of Neal (2003) is concerned with the sampling of $p(y\mid w)\propto {\bf 1}(\pi(y)>w)$ which is uniform, and let $S=\{y:\pi(y)>w\}$. The aim is to find an interval $I=(L,R)$ which contains the whole, or a part, of $S$, and to sample a proposal $y^*$ uniformly from $I$ and accept it as $y$ if $y^*\in S$. Now the interval $I$ will be constructed stochastically from $x=y_c$ and hence, as we are dealing with uniform densities; it is required that
$$p(y\mid x,w)=p(x\mid y,w).$$ 
Effectively, this boils down to the probability of getting $I$ from $x$ being the same as the probability of getting $I$ from $y$. Neal (2003) has two key ideas for constructing $I$ and we will focus on the ``stepping out'' procedure.

The idea here is to select a positive value $k$ and an integer $m\geq 1$ and start with 
$$L=x-k\,(1-U)\quad\mbox{and}\quad R=x+k\,U$$
where $U$ is a uniform random variable from $(0,1)$. It is already interesting to note that with $m=1$ this approach would coincide exactly with our own by choosing $s^{-1}p(s)$ to be a point mass of 1 at $s=k$.
This can be seen by noting that our algorithm selects $l$ uniformly from the interval $(x-k/2,x+k/2)$; i.e. 
$l=x-k/2+Uk$ and then takes $y^*$ uniformly from $(l-k/2,l+k/2)$ which can be written as
$(x-k(1-U),x+kU)$. 

To move on from this rather inflexible strategy, whereas with our algorithm we take $k=s$ as a random variable, Neal accounts for the rigidity of $k$ by allowing the interval to broaden out  by extending $L\to L-k$ and $R\to R+k$ until $\pi(L)<w$ and $\pi(R)<w$, respectively, or $J=0$ and $K=0$, respectively, where $J$ is a random number in $[0,\ldots,m-1]$ and $K=m-1-J$ and $J$ and $K$ go down by 1 every time an extension is made, respectively. The exact details are presented in Fig.~3 of Neal's paper where a proof is provided that this stochastic construction of $I$ does indeed satisfy detailed balance.

An alternative idea described in Neal (2003) is the ``doubling'' procedure and is described in Fig.~4 of his paper. 
The starting point is as with the stepping out procedure but now the intervals double in size when the interval is allowed to grow.
In short, the additional latent variables $l$ and $s$ we introduce at the outset obviate the need for a doubling or stepping out procedure.  So while we are able to treat $k=s$ as random within our framework, and hence deal with any issue arising as a consequence of it being fixed, it has recently been pointed out that some problems are sensitive to the choice of $k$ within Neal's slice sampler; see Karamanis and Beutler (2020).

\subsection{Numerical comparison}
We compared the latent slice sampler with the slice sampling algorithm by using the illustrations in section 8 of Neal's paper. It is a ten-dimensional funnel-like distribution of ten real-valued variables $v$ and $x_1$ to $x_9$. The marginal distribution of $v$ is Gaussian with mean zero and standard deviation 3. Conditional on a given value of $v$, the variables $x_1$ to $x_9$ are independent, with the conditional distribution for each being Gaussian with mean zero and variance $e^v$, which can be formulated as
$v \sim \text{N}(v\mid 0, 3^2)$ with $[x_i\mid v] \sim \text{N}(x_i\mid 0, e^v)$ for $i = 1, \dots, 9$.
The joint distribution is obviously given by
$$p(v, x_1,\dots, x_9) = \text{N}(v\mid 0, 3^2) \prod_{i = 1}^{9}\text{N}(x_i\mid 0, e^v).$$
Such a distribution is typical of priors for components of Bayesian hierarchical models; $x_1$ to $x_9$ might, for example, be random effects for nine subjects, with $v$ being the log of the variance of these random effects. If the data is largely informative, the problem of sampling from the posterior will be similar to that of sampling from the prior. From the above framework, we know the correct marginal distribution for $v$, which is the focus of the illustration, and we can sample for each of $x_1$ to $x_9$ given the value for $v$.

\setlength\belowcaptionskip{-3ex}
\begin{figure}[!htbp]
\begin{center}
\includegraphics[height=10cm,width=15cm]{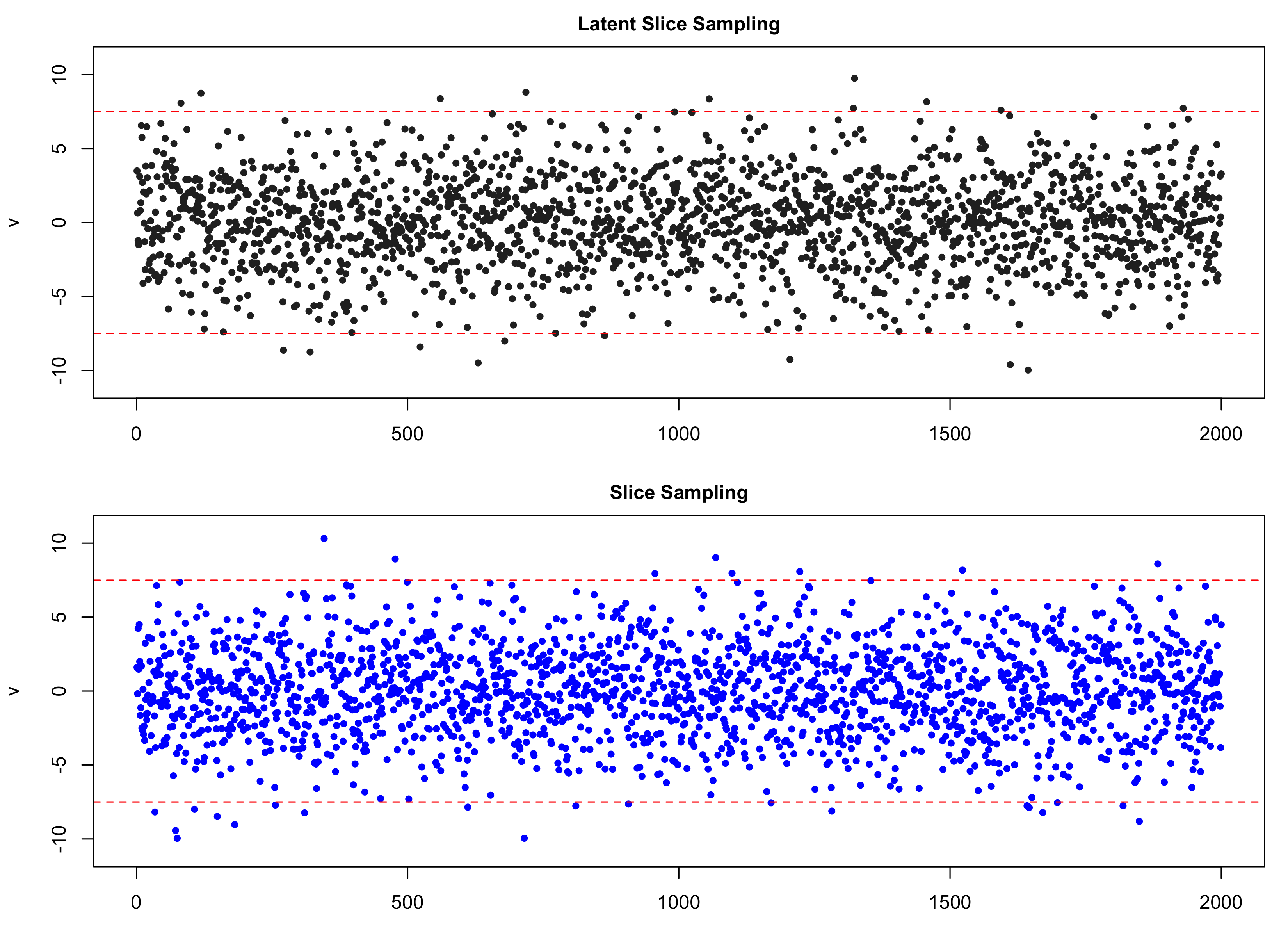}
\caption{Sampling the funnel distribution using latent slice sampling (dark dots) and single-variable slice sampling (blue dots)}
\label{funnel}
\end{center}
\end{figure}
 
In Neal's paper, the single variable slice sampling method is used to sample from a multivariate distribution by sampling repeated for each variable in turn. Each update uses the step-out and shrinkage procedure. Fig.~\ref{funnel} compared the result of trying to sample from the funnel distribution using latent slice sampling and single-variable slice sampling. The upper plot shows 2000 iterations of a run, which is the subsampling of 4,000,000 samples with a spacing of $m = 200$ to reduces the autocorrelation of successive samples. If every 200 th iteration is used and the rest thrown away, this produces another reversible Markov chain with asymptotic variance. The selection of spacing $m = 200$ can yield better estimates of the true posterior and yet smooth out autocorrelation. We use a gamma distribution with shape 2 and scale 5 to randomize the ``slice", i.e $p(s) \propto s e^{-s/5}$ so that the sampler is able to explore the distribution efficiently. The lower plot of Fig.~\ref{funnel} shows the results of trying to sample from the funnel distribution using single-variable slice sampling. To avoid the high autocorrelation, the same spacing of $m = 200$ is used to ``thin" the simulations. 

The resulting 2000 updates are shown in the scatterplot. Both the latent slice sampler and the single-variable slice sampling perform fairly well with small and large values of $v$ sampled quite good, compared with single-variable Metropolis updates and multivariate Metropolis updates, as discussed in Neal's paper. However, slice sampling method takes much greater cost in wasted computation. The average time for 10000 iterations are at least three times of that for latent slice sampling algorithm. The simplicity of the latent slice sampling makes it favorable for sampling distribution without selecting proposal distribution. By using stochastic search we accelerate the convergence to the stationary distribution.

\section{Illustrations}

In this section we present a number of illustrations. We start with two examples for discrete spaces, which include the allocation variables in a mixture of Dirichlet process model and the number of components in a mixture model.
We then consider some continuous examples, including a model in which Neal's slice sampler has been used, elliptical sampling, and then a state space model and a variable selection model where the vectors of unknowns are typically sampled componentwise using a Gibbs sampler. In these latter two examples we use the multivariate latent slice sampler to sample the entire vector as a single block.

\subsection{Mixture of Dirichlet process model}

Here we consider the well--known and widely used mixture of Dirichlet process (MDP) model, introduced in Lo (1984). The MDP model with Gaussian kernel is given by
$$f(x) = \int \text{N}(x \mid \mu, \sigma^2)\,d\,P(\theta)$$
where $\theta = (\mu, \sigma^2)$ and where $\mu$ represents the mean and $\sigma^2$ the variance of the normal kernel. Let $\text{DP}(\alpha, P_0)$ denote a Dirichlet process prior (Ferguson, 1973) with scale parameter $\alpha > 0$ and a prior probability $P_0$, so $\mbox{E}(P)=P_0$ and $\mbox{Var}(P(A)) = P_0(A)\,(1-P_0(A)/(1+\alpha)$ for all appropriate sets $A$. The model has been one of the most popular in Bayesian nonparametrics, and for a recent review on estimation techniques, see Hjort et al (2010).

A number of recent ideas maintain the $P$ as part of a Markov chain sampling approach and find ways to obviate the need for sampling an infinite dimensional object. The more traditional approaches marginalize it out of the model. Here we adopt the former approach and the key idea here is to introduce the latent indicator variable which tells us which component each observation came from. Following Sethuraman (1994) we can write
$$P=\sum_{j=1}^\infty w_j\,\delta_{\theta_j}$$
where $w_1=v_1$ and $w_j=v_j\prod_{l<j}(1-v_l)$ with the $(v_j)$ i.i.d. $\mbox{beta}(1,\alpha)$, and the $(\theta_j)$ are i.i.d. $P_0$. If we now let $d_i\in\{1,2,\ldots\}$ indicate the component number of $x_i$, the complete likelihood function is given by
 \begin{align}
     l(x, d\mid w, \theta) = \prod_{i=1}^n w_{d_i}\text{N}(x_i\mid\theta_{d_i})\nonumber
 \end{align}
To complete the prior set up we determine $P_0$. The prior for the $(\mu_j)$ is independent $\text{N}(0, 1/s)$ and the prior for the  $(\lambda_j)=1/\sigma^2_j$ will be independent $\text{gamma}(\tau,\tau)$. The full conditional distributions for the parameters $(\mu,\lambda,v)$ are standard, being independent normal, gamma, and beta respectively, given the data and the $(d_i)$. We omit the details as they are well documented in the literature. See, for example, Walker (2007).

On the other hand, the conditionals for the $(d_i)$ are difficult due to the fact that the normalizing constant is not available. So
$$P(d_i=j\mid\cdots)=\pi(j)\propto w_j\,\text{N}(x_i\mid\mu_j,\sigma^2_j).$$
Rather than attempting to sample this directly, which is impossible, we use the transition density (\ref{discrete}) with a choice of finite $k$; so, with $d_i$ denoting the current value and $d_i'$ the new to be sampled value,
$$P(d_i' = j\mid d_i, \cdots) = \frac{\pi(j)}{k}\sum_{l=\max(j,\,d_i)}^{\min(j+k-1,\,d_i+k-1)}\frac{1}{\sum_{z=\max(1,\,l-k+1)}^l \pi(z)},$$
with  $|j - d_i| <k $.

The infinite dimensional problem automatically converts to a finite one but which retains a valid Markov chain with the correct stationary distribution. On the other hand, Ishwaran and James (2010) truncated $\pi(j)$ to some large value which obviously introduces errors.  

Our aim here is not to undertake an extensive simulation exercise or wide ranging comparison, but rather to demonstrate the simplicity and accuracy of using this density for sampling the  $(d_i)$.  The point is that the stationary distribution is correct while no extra latent variables are being introduced to sample the $(d_i)$. On the other hand, a number of extra latent variables were required in Kalli et al  (2011). The simulation for the MDP model is a normal example of 400 random variables sampled independently from the density $f(x) = \frac{1}{3}\text{N}(x\mid-4,1) + \frac{1}{3}\text{N}(x\mid 0,1) + \frac{1}{3}\text{N}(x\mid8,1)$. For illustrative purposes, we took $\tau = 0.5$, $s = 1$, and $\alpha = 2$. The Gibbs sampler was run for 20,000 iterations and at each iteration from 15,000 onwards a predictive sample $x_{n+1}$ was taken. 

\setlength\belowcaptionskip{-3ex}
\begin{figure}[!htbp]
\begin{center}
\includegraphics[height=8cm,width=14cm]{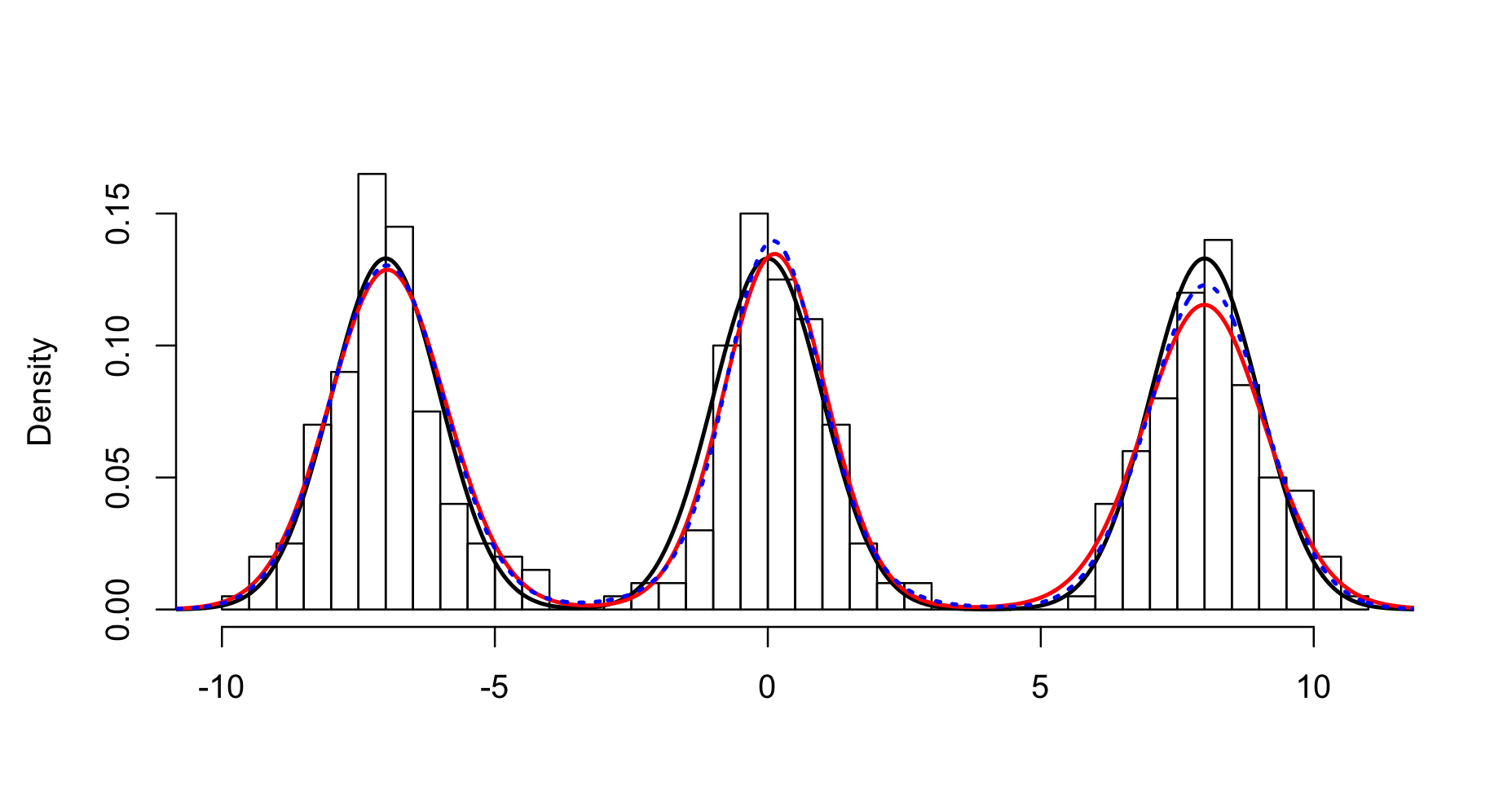}
\caption{Histogram of the data and predictive densities from chain using new transition density (red) and slice--efficient sampler (blue dotted) and the true density (black)}
\label{mdp}
\end{center}
\end{figure}

A histogram of the 400 data points with the density estimators (blue: using our new transition density approach; red: using slice sampling as described in Kalli et al (2011)) based on 5000 samples of $x_{n+1}$, and with the true density (black), are provided in Fig.~\ref{mdp}. The density estimators were obtained using the R density routine from the predictive samples. The advantage of using our new transition density is that we do not need any truncation of the distribution of the $(d_i)$. After picking an appropriate $k$, there are no other parameters to be tuned and the algorithm itself is straightforward. It avoids the accept/reject component of a Metropolis--Hastings algorithm, the errors introduced by truncating the correct density of the $(d_i)$, and avoids introducing further latent variables to sample the $(d_i)$.

\subsection{Mixture model: unknown number of components}

Here we consider another mixture model but now we use a version which works with a random number of components. For illustrative purposes we select a case where the components are fully specified, exponential densities with the integers as parameter. A more complete version of the model with unknown normal components was considered in Richardson and Green (1997).

The model, given  $M$, the number of components, is given by
$$f(x\mid w_M, M) = \sum_{j=1}^M w_{jM}j e^{-jx}$$
with $M\in \{1,\ldots\infty\}$ and $w_M=(w_{1M}, \dots, w_{MM})$ are the weights which sum to 1. 
Using the indicator variables $(d_i)$, as in the previous section, though now given $M$ their values will be from a finite set, the complete likelihood function is given by
$$l(w, M\mid x,d) = \pi(w\mid M)\,\pi(M)\prod_{i=1}^n w_{d_i}d_i e^{-d_i\,x_i}.$$
Here $w$ represents all weights for all possible $M$; i.e. $w=(w_1,w_2,\ldots)$. We adopt the framework of Godsill (2001) and provide a prior for each $w_j$ given $M$, in the form
$$\pi(w\mid M)=\pi(w_2\mid w_3)\ldots\pi(w_{M-1}\mid w_M)\pi(w_M\mid M)\pi(w_{M+1}\mid w_M)\pi(w_{M+2}\mid w_{M+1})\ldots\,.$$  
The prior for $M$ is
$\pi(M) = \lambda^{M-1}e^{-\lambda}/(M-1)!$ for $M=1,2\ldots$, so is a Poisson shifted to $\{1,2,\dots\}$.
The prior for $w_M$ given $M$ is Dirichlet with common parameter $\alpha$. To complete the prior setting, we need to specify $\pi(w_{j+1}\mid w_j)$ and $\pi(w_{j-1}\mid w_j)$, the latter avoiding $j=1$. The former is obtained by selecting a weight from $(w_{jl})_{l=1:j}$ and splitting it into two, $uw_j$ and $(1-u)w_j$, with $u$ uniform on $(0,1)$. Hence $\pi(w_{j+1}\mid w_j)=1/j$. Likewise, for $\pi(w_{j-1}\mid w_j)$ we take $w_{jj}$ and combine it with $w_{jl}$, for $l\ne j$. Hence, $\pi(w_{j-1}\mid w_j)=1/(j-1)$.

For the ensuing Gibbs sampler Markov chain, all the full conditionals are easy to sample, including $w_M$ given $M$ and the $(d_i)$ given $M$. However, sampling $M$ is the difficulty. The benchmark procedure here is a reversible jump Markov chain, see Green (1995), where a detailed balance condition is required. While this may not be difficult for the present mixture of known exponential components, it is far from trivial for unknown normal components; see Richardson and Green (1997). Now 
$$\pi(M\mid \cdots) \propto \pi(M)\,\pi(w_M\mid M) \prod_{i=1}^n \sum_{j=1}^M w_{jM} j e^{-jx_i} 
    \times \prod_{j=2}^{M-1}\pi(w_j\mid w_{j+1})\,\prod_{j=M+1}^\infty \pi(w_j\mid w_{j-1}).$$
It is important to note that for any $M'\ne M$,
$$\frac{\pi(M'\mid \cdots)}{\pi(M
\mid \cdots)}=\frac{\pi(M')\,\pi(w_{M'}\mid M') \prod_{i=1}^n \sum_{j=1}^{M'} w_{jM'} j e^{-jx_i} }{\pi(M)\,\pi(w_M\mid M) \prod_{i=1}^n \sum_{j=1}^M w_{jM} j e^{-jx_i} };$$
i.e. because of how we set the conditional priors for the weights, they cancel from the ratio of posteriors for different number of component values.     
So we now sample the new $M'$ given the current $M$, with a chosen $k$, via the transition density
$$P(M'\mid M) = \frac{\pi(M')}{k} \sum_{l = \max(M',\, M)}^{\min(M' + k - 1,\,M + k - 1)}\frac{1}{\sum_{z = \max(1,\,l - k + 1)}^l \pi(z)},$$
with $|M'-M|<k$.

A reversible jump Markov chain from a current $M$ would typically propose a move to either $M-1$ or $M+1$ and up front sample a set of weights which the chain would move to if the proposed move was accepted. As with all such algorithms, it has an accept/reject component which, to reiterate, our transition density does not have.

\setlength\belowcaptionskip{-3ex}
\begin{figure}[!htbp]
\begin{center}
\includegraphics[height=7.5cm,width=14cm]{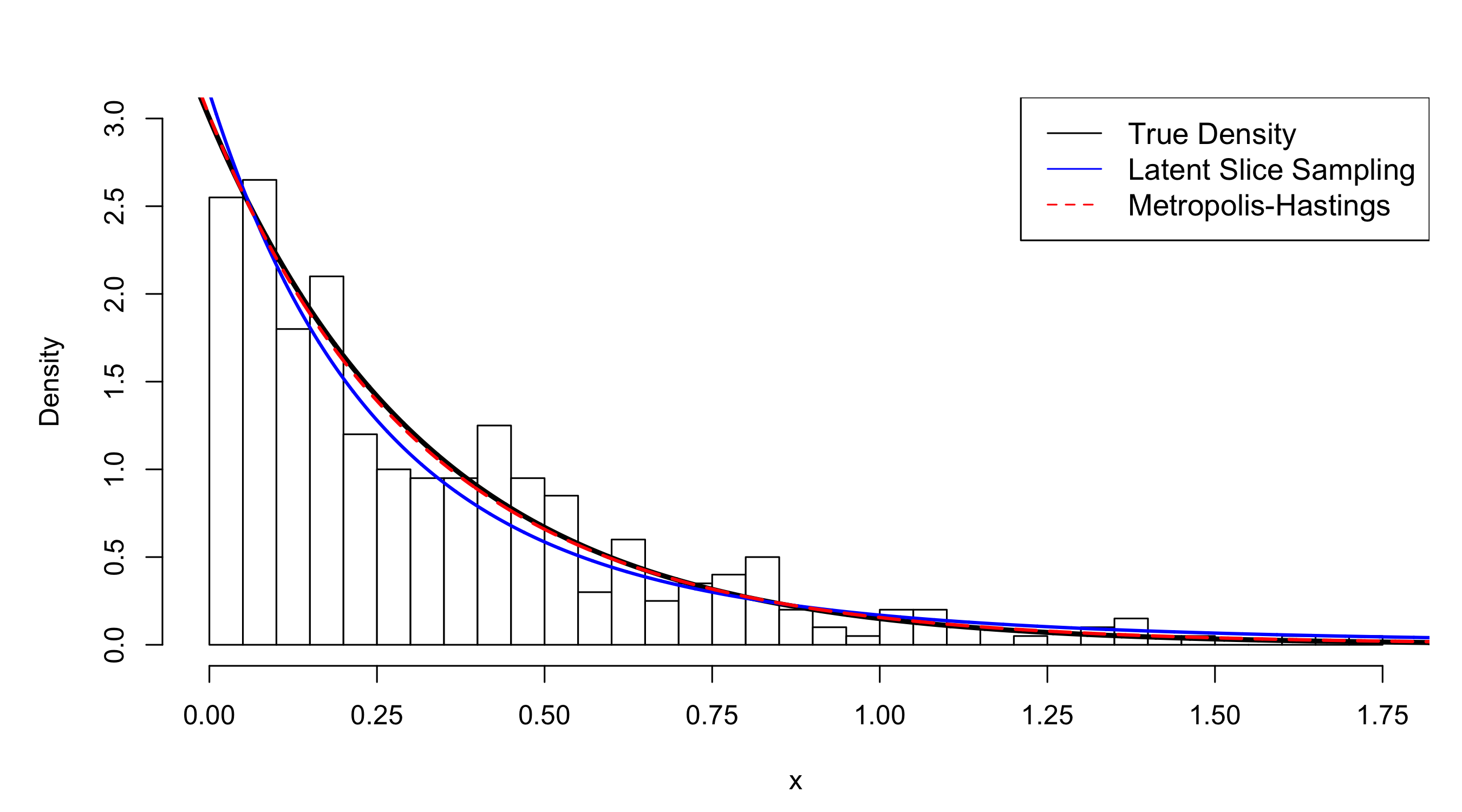}
\caption{Predictive densities using new transition density and reversible jump algorithms, with histogram of data.}
\label{mfm}
\end{center}
\end{figure}

For the demonstration we generated 400 i.i.d. data points from a single exponential density with parameter $3$, i.e. $f(x) = 3e^{-3x}.$ Fig.~\ref{mfm} shows the histogram of the data along with the predictive density estimates using the new transition density and show alongside the estimate from the reversible jump algorithm. Both are clearly working well.


\subsection{Elliptical sampling}

In this example we consider a continuous space, and here we label the algorithm with the new transition density as the latent slice sampler.
Specifically, here,  we compare with elliptical slice sampling, which is used in a number of models which have a  multivariate Gaussian distribution as the prior. See Murray et al (2010). The objective is to sample from a posterior distribution over latent variables that is proportional to the product of a multivariate Gaussian prior and a likelihood function that ties the latent variables to the observed data. 

Suppose $\mathbf{f}$ is the vector of the latent variables that we wish to sample and has a zero--mean multivariate Gaussian prior with covariance matrix $\Sigma$; i.e. $\mathbf{f} \sim \text{N}(\mathbf{0}, \Sigma)$ and, for completeness, the density function is  given by
$$\text{N}(\mathbf{f}\mid \mathbf{0}, \Sigma) \equiv |2\pi\,\Sigma|^{-1/2}\text{exp}\left(-\half\,\mathbf{f}^{\text{T}}\,\Sigma^{-1}\,\mathbf{f}\right).$$
The data are assume to have likelihood function $L(\mathbf{f}) = p(\text{data}\mid \mathbf{f})$ so that the target posterior  distribution  is
$$\pi^*(\mathbf{f}) \propto \text{N}(\mathbf{f}\mid \mathbf{0}, \Sigma)\,\,L(\mathbf{f}).$$
Given a current state $\mathbf{f}$, a new state can be proposed via
$$\mathbf{f}' = \sqrt{1-\epsilon^2}\, \mathbf{f} + \epsilon\, \boldsymbol{\nu}\quad\mbox{with}\quad \boldsymbol{\nu} \sim \text{N}(\mathbf{0}, \Sigma),$$
where $\epsilon\in[-1, 1]$ is a step-size parameter, and the proposal is accepted or rejected using a Metropolis--Hastings step. However, apparently the choice of $\epsilon$ becomes crucial. A more flexible approach would allow for a richer class of proposal. An alternative and more natural  parameterization for the proposal is 
$$\mathbf{f}' = \boldsymbol{\nu}\sin\theta + \mathbf{f}\cos\theta,$$
defining a full ellipse as $\theta$ ranges over $[0,2\pi]$.
The original strategy in Murray et al (2010) is to take $\theta$ as random and sample $\theta$ using Neal's slice sampler; here we replace this part with our latent slice sampler. The resulting algorithm is given in Table 1.

\begin{table}
\caption{Modified elliptical slice sampling with latent slice algorithm.}
\begin{tabular}{l} 
 \hline
 \textbf{Input}: current state $\mathbf{f}$, log-likelihood function $\log L$ \\ 
 \multicolumn{1}{p{17cm}}{\textbf{Output}: new state $\mathbf{f}'$. }\\ \hline
 1. Sample $\boldsymbol{\nu}\sim \text{N}(\mathbf{0}, \Sigma)$. \\
 2. Sample $u\sim \text{U}(0,1)$ and set $\log w \leftarrow \log L(\mathbf{f}) + \log u$. \\  
 3. Sample $\theta$ using latent slice sampler. \\
 4. $\mathbf{f}' \leftarrow \mathbf{f}\cos \theta+\boldsymbol{\nu}\sin\theta$: if $\log L(\mathbf{f}') > \log w$, accept  $\mathbf{f}'$; else GoTo step 3. \\
 \hline
\end{tabular}
\end{table}

For our illustration we consider a Gaussian data model; i.e.
$$y_i=f(x_i)+\varepsilon_i\quad\mbox{with}\quad \varepsilon_i\sim \text{N}(0,\sigma^2),$$
assuming $\sigma$ is known. The Gaussian process prior has covariance matrix given by elements
$$\Sigma_{i,j} = \tau^2\exp \left( -\frac{\sum(x_{i}-x_{j})^2}{2\psi^2}\right),$$
where $\psi$ is the ``lengthscale" parameter and $\tau$  the  ``signal variance". 
So we take
$$L(\mathbf{f}) \propto \exp\left\{-\half \sum_{i=1}^n (y_i-f(x_i))^2\right\}.$$

In the experiment, we generate a sequence of $n=100$ evenly spaced values $(x_{1:n})$ over the interval $[0, 1]$ as the input data and take the true function $f(x_i) = \sin(4\pi\,x_i) + \sin(7\pi \, x_i)$. We take the noise standard deviation as $\sigma=0.2$ to generate the data $(y_{1:n})$. For the covariance matrix of the Gaussian process prior, we use lengthscale $\psi=0.1$ and unit signal variance, $\tau=1$.

\setlength\belowcaptionskip{-3ex}
\begin{figure}[!htbp]
\begin{center}
\includegraphics[height=7cm,width=16cm]{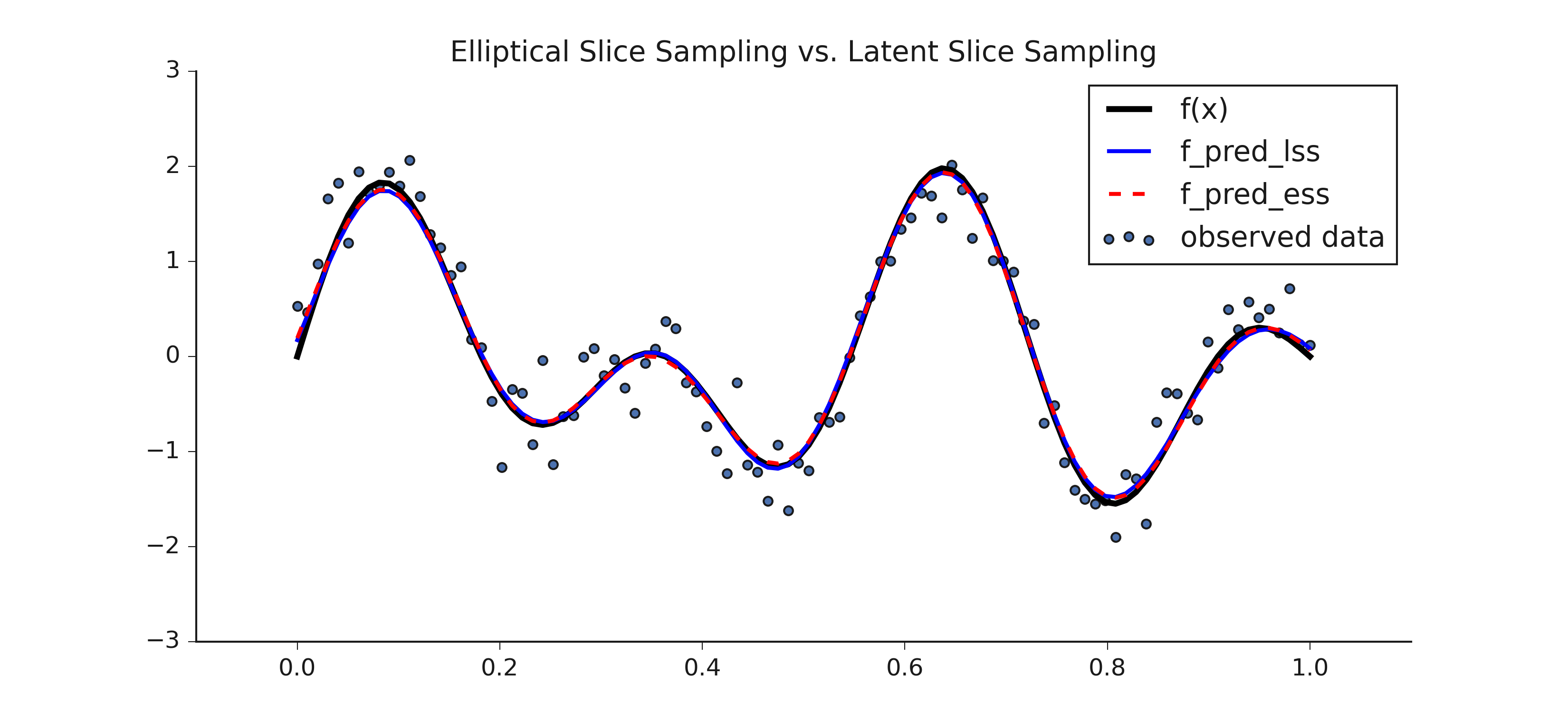}
\caption{Comparison of estimated latent function from latent slice sampling(blue) and elliptical slice sampling(red dash). The black solid curve is the observed latent function.}
\label{elliptical}
\end{center}
\end{figure}

Fig.~\ref{elliptical} shows the estimated function using our own latent slice sampler and also Neal's slice sampler. Both, obviously, perform well with fast convergence, indicating that the latent slice sampler can be applied in a vast range of Gaussian based models that are currently using Gibbs, Metropolis--Hastings, or slice sampling.

\subsection{State space model} In this subsection we sample a 500 dimensional space which is the unknown states of a state space, also known as a hidden Markov, model. We consider 
$$[y_i\mid x_i]\sim \mbox{Poisson} \left(\theta \,\exp(x_i)\right) \quad\mbox{and}\quad x_i=\rho \, x_{i-1}+\sigma \, z_i$$
for $i=1,\ldots,n$ with $n=500$ and $x_0=0$ and the $(z_i)$ independent standard normal. To generate the data set we take $\rho=0.8$, $\sigma=1$ and $\theta=1$. 

The joint density of the $x=(x_{1:n})$ given $\theta$ is
$$\pi(x\mid\theta)\propto \exp\left\{\sum_{i=1}^n \left[x_i\,y_i -\theta\, e^{x_i} - \half(x_i-\rho\, x_{i-1})^2\right] \right\};$$
for simplicity we assume $\rho$ and $\sigma$ to be known, without any loss to the illustration about to be presented. 
Typically, the $\pi(x\mid\theta)$ is sampled component by component, i.e. by sampling $p(x_i\mid x_{-i},\theta)$
for $i=1,\ldots,n$ within a Gibbs sampling framework. In some special cases, conditionally normal dynamic linear models, it can be sampled as a block by backward sampling. The most common approaches nowadays are based on particle filters; see Andrieu et al. (2010).

Using the multivariate latent slice sampling algorithm we sample the entire vector of state spaces in one block.
We only assume $\theta$ is unknown and the conditional density of $\theta$ with a gamma prior with shape and rate parameters both equal to $0.5$ is given by a gamma distribution with shape parameter $0.5+\sum_{i=1:n} y_i$ and rate parameter $0.5+\sum_{i=1:n}e^{x_i}$.

\begin{figure}[!htbp]
\begin{center}
\includegraphics[height=8cm,width=14cm]{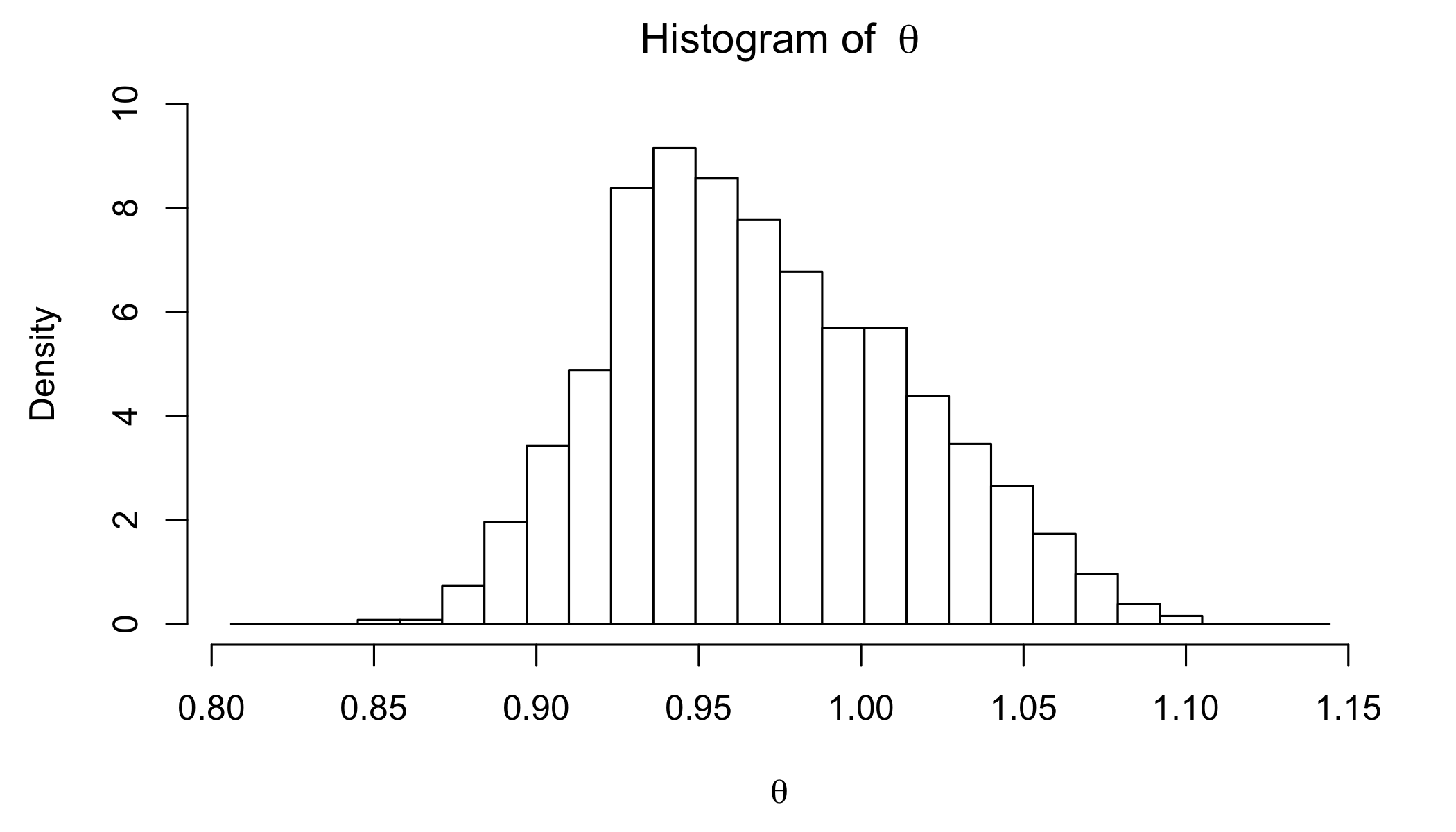}
\caption{Posterior density of $\theta$ for state space model}
\label{figst}
\end{center}
\end{figure}

The chain was run for 2000 iterations and the time taken was 20 secs. A plot of the posterior $\theta$ samples is presented in Fig.~\ref{figst}. The mean value is 0.97.


\subsection{Spike and slab model} In this subsection we consider a popular approach to variable selection within the Bayesian framework; namely the spike and slab prior (George and McCulloch, 1993).
The model is given by
$$Y = X\beta + \epsilon, \quad \epsilon\sim \mbox{N}(0, \sigma^2\mathbf{I}_n)$$
where $Y\in \mathbb{R}^n$ is a vector of responses, $X=[X_1, \dots, X_p] \in \mathbb{R}^{n\times p}$ is a regression matrix of $p$ predictors, $\beta=(\beta_1,\ldots,\beta_p)^T \in \mathbb{R}^p$ is a vector of unknown regression coefficients, and $\epsilon \in \mathbb{R}^n$ is the noise vector of independent normal random variables with $\sigma^2$ as their unknown common variance. The spike and slab prior for $\beta$ is given by
$$\pi(\beta)\propto\prod_{j=1}^p\left[\sigma_1^{-1}\exp(-\half\beta_j^2/\sigma_1^2)+\sigma_2^{-1}\exp(-\half\beta_j^2/\sigma_2^2)\right],$$
where $\sigma_1\approx 0$ yields the spike and $\sigma_2\approx\infty$ yields the slab. Markov chain Monte Carlo methods for this model require the Gibbs sampling of $\beta_j$ conditional on the $\beta_{-j}$, i.e. the vector of $\beta$ without the $\beta_j$. See, for example, Narisetty and Xe (2014).
Here we use the latent slice sampler to sample $\beta$ as one block.

We assume $\sigma=1$ is known and generate data for $n=100$ with $p=90$. We take $\beta_1=1$, $\beta_{2:5}=5$ and $\beta_{6:90}=0$. All the elements in the design matrix $X$ are generated as independent standard normal random variables. We take $\sigma_1=0.1$ and $\sigma_2=10$; writing down the posterior for $\beta$ is quite straightforward and is in particular easy to compute for any given value of $\beta$. We ran the latent slice sampler for  10,000 iterations; taking a few seconds to complete the task.

\begin{figure}[!htbp]
\begin{center}
\includegraphics[height=6.5cm,width=14cm]{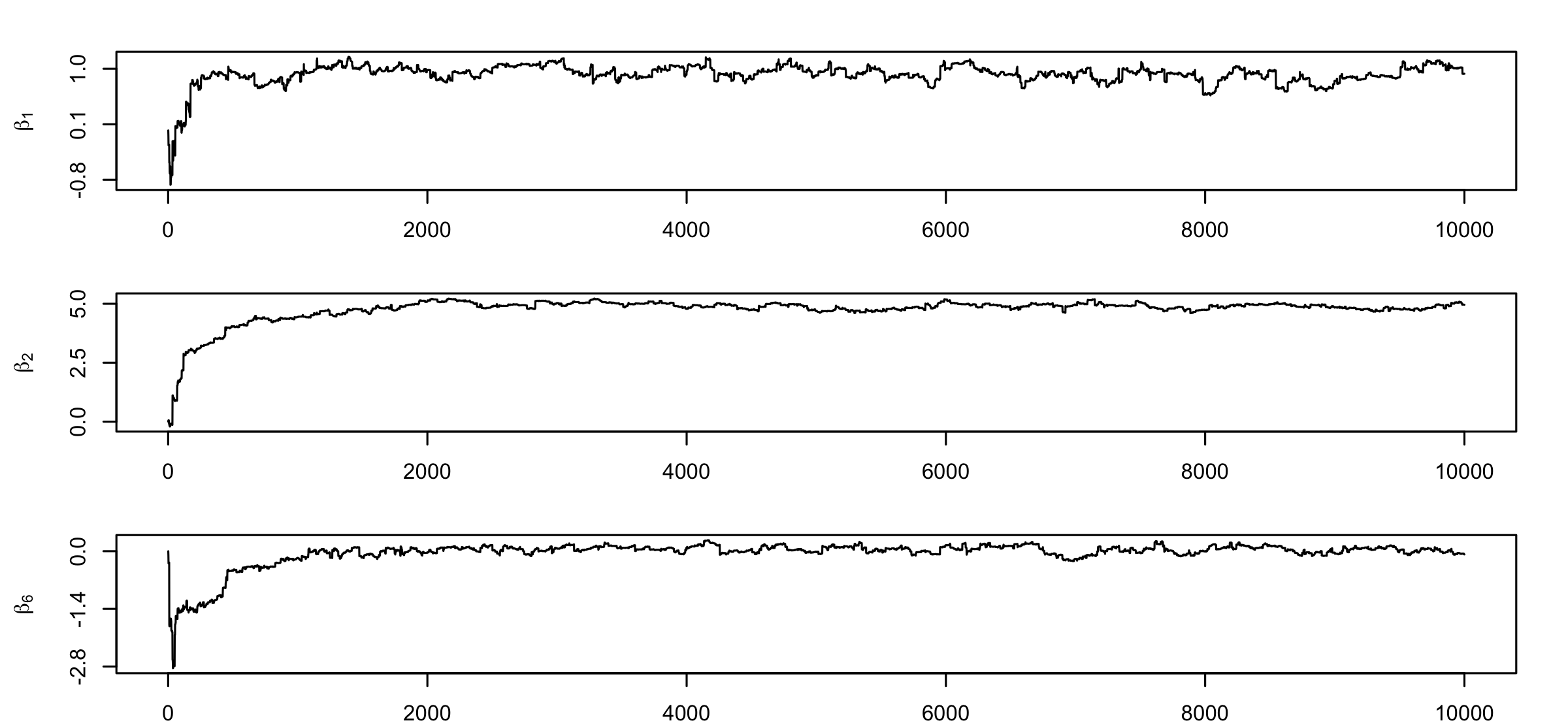}
\caption{Posterior samples of $\beta_1$, $\beta_2$ and $\beta_6$ from spike and slab model}
\label{figssp}
\end{center}
\end{figure}

For illustration we present the posterior samples for $\beta_1$ and $\beta_2$ and $\beta_6$; the true values being $1$, $5$ and $0$, respectively. As is visible from Fig.~\ref{figssp} the samples are accumulating at the correct locations.

\section{Discussion} In this paper we have presented a generic sampling algorithm which has the ability to sample efficiently very high dimensional distribution functions at great speed. The key is the latent model combined with the shrinkage procedure based on uniform distributions and an automatic reversible condition. Given the simplicity of the algorithm we present it here, in the general $d$--dimensional case, with target density $\pi(y)$ and $y=(y_1,\ldots,y_d)$. Let $\lambda=0.1$, for example; we describe a single loop with current values $y_0=(y_{01},\ldots,y_{0d})$ and $s_0=(s_{01},\ldots,s_{0d})$.
 
\begin{itemize}
    \item[1.] Sample $w\sim \mbox{U}(0,\pi(y_0))$ and, for $j=1,\ldots,d$, sample
    $$l_{j}\sim \mbox{U}\big(y_{0j}-s_{0j}/2,y_{0j}+s_{0j}/2\big)$$
    and sample $s_j$ from the density proportional to
    $$\exp(-\lambda s_j)\,{\bf 1}(s_j>2|l_j-y_{0j}|).$$
    
    \item[2.] Set $a_j = l_j-s_j/2$ and $b_j = l_j+s_j/2$. 
    
    \item[3.] For $j=1,\dots,d$, sample
    $$y_j^{*} \sim \mbox{U}(a_j,b_j).$$
    
    if $\pi(y^{*})>w$, accept $y=y^{*}$; else, for $j=1,\ldots,d$,
    $$\mbox{if}\quad y_j^{*}<y_{0j}\quad\mbox{then}\quad  a_j\leftarrow \max\{a_j,y_j^{*}\}\quad\mbox{else}
    \quad b_j\leftarrow \min\{b_j,y_j^{*}\}.$$
    
    \item[4.] Repeat step 3 until $\pi(y^{*})>w$ and set $y=y^{*}$.
 
\end{itemize}
\noindent
As we have demonstrated, such an algorithm can work with a nonlinear state space model with dimension 500 and return output in short time. Future work will consider sampling of constrained spaces, such as uniform sampling on polytopes and truncated distributions, such as the multivariate normal (Robert, 1995; Damien and Walker,  2001).

\section*{References}

\begin{description}

\item Andrieu, A., Doucet, A. and Holenstein, R. (2010). Particle Markov chain Monte Carlo methods. {\em Journal of the Royal Statistical Society, Series B} {\bf 72}, 269--342.

\item Besag, J. and Green, P.~J. (1993). Spatial statistics and Bayesian computation. {\em Journal of the Royal Statistical Society, Series B} {\bf 55}, 25--37. 

\item Damien, P., Wakefield, J.~C. and Walker, S.~G. (1999). Gibbs
sampling for Bayesian nonconjugate and hierarchical models using
auxiliary variables. {\em Journal of the Royal Statistical
Society, Series B} {\bf 61}, 331--344.

\item Damien, P. and Walker, S.G. (2001). Sampling truncated normal, beta and gamma densities. {\em Journal of Computational and Graphical Statistics} {\bf 10}, 206--215. 

\item Ekin, T., Walker, S.~G. and Dmaien, P. (2020). Augmented simulation methods for discrete stochastic optimization with recourse. To appear in {\em Annals of Operations Research}.

\item Ferguson, T.~S. (1973). A Bayesian Analysis of Some Nonparametric Problems. {\em The Annals of Statistics} {\bf 1}, 209--230 . 

\item George, E.~I. and McCulloch, R. (1993). Variable selection via Gibbs sampling. {\em Journal of the American Statistical Association} {\bf 88}, 881--889.

\item Godsill, S.~J. (2001). On the relationship between MCMC methods for model uncertainty. {\em Journal of Computational and Graphical Statistics} {\bf 10}, 230--248.  

\item Green, P.~J. (1995). Reversible jump MCMC computation and Bayesian model dtermination. {\em Biometrika} {\bf 82}, 711--732.

\item Hastings, W.~K. (1970). Monte Carlo sampling methods using Markov chains and their applications. {\em Biometrika} {\bf 57}, 97--109.

\item Hjort, N.~L., Holmes, C., Mueller, P. and Walker, S.~G. (2010). {\em Bayesian Nonparametrics}. Cambridge University Press.

\item Ishwaran, H. and James, L.~F. (2001). Gibbs sampling methods for stick--breaking priors. {\em Journal of the American Statistical Association} {\bf 96},  161--173.

\item Kalli, M., Griffin, J.~E. and Walker, S.~G. (2009).  Slice sampling mixture models. {\em Statistics \& Computing} {\bf 21}, 93--105.

\item Karamanis, M. and Beutler, F. (2020). Ensemble slice sampling. ArXiv:2002.06212v1.

\item Lo, A.~Y. (1984). On a class of Bayesian nonparametric estimates: I. Density estimates. {\em Annals of Statistics}, {\bf 12}, 351--357.  

\item Metropolis, N., Rosenbluth, A.~W., Rosenbluth, M.~N., Teller, A.~H. and Teller, E. (1953).
Equation of state calculations by fast computing machines. {\em The Journal of Chemical Physics} {\bf 21},  1087--1092.

\item Mira, A. and Tierney, L. (2002). Efficiency and convergence properties of slice samplers. {\em Scandinavian
Journal of Statistics} {\bf 29}, 1--12.

\item Murray, I., Adams, R.~P. and Mackay, D.~J.~C. (2010). Elliptical slice sampling. {\em Journal of Machine Learning Research} {\bf 9}, 541--548.  

\item Narisetty, N.~N. and He, X. (2014). Bayesian variable selection with shrinking and diffusing priors. {\em Annals of Statistics} {\bf 42}, 789--817.

\item Neal, R.~M. (2003). Slice sampling. {\em Annals of Statistics} {\bf 31}, 705--767.

\item Nishihara, R., Murray, I. and Adams, R.~P. (2014). Parallel MCMC with generalized ellipitcal slice sampling. {\em Journal of Machine Learning Research} {\bf 15}, 2087--2112.

\item Richardson, S. and Green, P.~J. (1997). On Bayesian analysis of mixtures with an unknown number of components. {\em Journal of the Royal Statistical Society, Series B} {\bf 59}, 731--792. 

\item Robert, C.~P. (1995). Simulation of truncated normal variables. {\em Statistics \& Computing} {\bf 5}, 121--125.

\item Roberts, G.~O. and Rosenthal, J.~S. (1999). Convergence of slice sampler Markov chains.
{\em Journal of the  Royal Statistical Society, Series B}  {\bf 61}, 643--660.

\item Sethuraman, J. (1994). A constructive definition of Dirichlet priors. {\em Statistica Sinica} {\bf 4}, 639--650.

\item Tibbits, M.~M., Haran, M. and Liechty, J.~C. (2011). Parallel mulitvariate slice sampling. {\em Statistics and Computing} {\bf 21}, 415--430.

\item Tibbits, M.~M., Groendyke, C., Haran, M. and Liechty, J.~C. (2014). Factor slice sampling. {\em Journal of Computational and Graphical Statistics} {\bf 23}, 543--563.

\item Walker, S.~G. (2007). Sampling the Dirichlet mixture model with slices. {\em Communications in Statistics} {\bf 36}, 45--54.

\item Walker, S.~G. (2014). Sampling un--normalized probabilities: An alternative to the Metropolis--Hastings algorithm. {\em SIAM Journal on Scientific Computing} {\bf 36}, A482--A494.

\end{description} 



\end{document}